\documentstyle[12pt,epsf]{article}
\renewcommand{\baselinestretch}{1.0}

\newcommand{\be}{\begin{equation}}
\newcommand{\ee}{\end{equation}}
\begin{document}
\topmargin 0pt
\oddsidemargin=-0.4truecm
\evensidemargin=-0.4truecm
\renewcommand{\thefootnote}{\fnsymbol{footnote}}

\newpage
\setcounter{page}{1}
\begin{titlepage}     
\vspace*{-2.0cm}
\begin{flushright}
\vspace*{-0.2cm}
\end{flushright}
\vspace*{0.5cm}

\begin{center}
{\Large \bf Solar neutrinos: the SNO salt phase results and physics of
conversion}
\vspace{0.5cm}

{P. C. de Holanda$^{1}$ and A. Yu. Smirnov$^{2,3}$\\
\vspace*{0.2cm}
{\em (1) Instituto de F\'\i sica Gleb Wataghin - UNICAMP, 13083-970
Campinas SP, Brasil}\\ {\em (2) The Abdus Salam International Centre
for Theoretical Physics, I-34100 Trieste, Italy }\\ {\em (3) Institute
for Nuclear Research of Russian Academy of Sciences, Moscow 117312,
Russia}

}
\end{center}

\begin{abstract}
We have performed analysis of the solar neutrino data including
results from the SNO salt phase as well as the combined analysis of
the solar and the KamLAND results.  The best fit values of neutrino
parameters are $\Delta m^2 = 7.1 \cdot 10^{-5}$ eV$^2$, $\tan^2 \theta
= 0.40$ with the boron flux $f_B = 1.04$.  New SNO results strongly
disfavor maximal mixing and the h-LMA region ($\Delta m^2 > 10^{-4}$
eV$^2$) which is accepted now at the $3\sigma$ level.  We find the
$3\sigma$ upper bounds: $\Delta m^2 < 1.7 \cdot 10^{-4}$ eV$^2$ and
$\tan^2 \theta < 0.64$, and the lower bound $\Delta m^2 > 4.8 \cdot
10^{-5}$ eV$^2$.  Non-zero 13-mixing does not change these results
significantly.  The present data determine quantitatively the physical
picture of the solar neutrino conversion. At high energies relevant
for SNO and Super-Kamiokande the deviation of the effective survival
probability from the non-oscillatory value is about 10 - 14\%.  The
oscillation effect contribution to this difference is about 10\% and the
Earth regeneration about 3 - 4 \%. At low energies ($E < 1$ MeV) the
matter corrections to vacuum oscillation effect are below 5\%.  The
predictions for the forthcoming measurements are given which include
the spectral distortion and CC/NC ratio at SNO, the Day-Night
asymmetry, the KamLAND spectrum and rate.

\end{abstract}

\end{titlepage}
\renewcommand{\thefootnote}{\arabic{footnote}}
\setcounter{footnote}{0}
\renewcommand{\baselinestretch}{0.9}

\section{Introduction} 

The SNO-II (salt phase) results \cite{salt} have further confirmed
correctness of both the neutrino fluxes predicted by the Standard
Solar Model (SSM)~\cite{ssm} and picture of the solar neutrino
conversion based on the MSW effect~\cite{w,ms}.  Together with the
results from the SNO phase-I~\cite{sno1}, Homestake~\cite{Cl},
SAGE~\cite{sage}, GALLEX~\cite{gallex}, GNO~\cite{gno} and
Super-Kamiokande~\cite{SK,SK2} experiments as well as from the reactor
experiment KamLAND~\cite{kamf} the latest SNO results lead to better
determination of the oscillation parameters.  This improvement allows
one to make two important qualitative statements~\cite{salt}:

\begin{itemize}

\item
the h-LMA region, which corresponds to $\Delta m^2 > 10^{-4}$ eV$^2$,
is strongly disfavored (being accepted at the $3\sigma$ level);

\item 
there is a substantial deviation of 1-2 mixing from the maximal one,
and the latter is rejected by more than $5\sigma$ standard deviations.

\end{itemize}

These statements have been confirmed by followed
analyzes~\cite{balan,fogli,valle2,alia,crem,choubey}.

There are two key points of the SNO-II publication: (i) new
measurement of the total active Boron neutrino flux with enhanced
neutral current sensitivity, and (ii) data analysis without assumption
of the undistorted neutrino spectrum.  They lead to larger (than
pre-salt) value of the neutrino flux measured by the neutral current
(NC) reaction and smaller flux measured by the charged current (CC)
reaction~\cite{salt}.  As a result, the ratio of neutrino
fluxes~\cite{salt}:
\be
\frac{\rm CC}{\rm NC} = 0.306 \pm 0.026(stat) \pm 0.024(syst) 
\label{cc/nc}
\ee
turns out to be smaller than the value from pre-salt measurements:
$0.346 + 0.048/-0.046$ \cite{sno1}. It is, however, larger than the
value which corresponds to the NC flux determined previously without
constraint of the undistorted spectrum: CC/NC $ = 0.274 \pm 0.073$
\cite{sno1}.

The ratio (\ref{cc/nc}) is also smaller than the expected one.  In the
previous paper~\cite{ped02b} we have predicted that the central value
of the ratio (which corresponds to the best fit point) and the
$3\sigma$ error bars equal: CC/NC $= 0.33 + 0.15/-0.07$. It was
pointed out that values CC/NC$ < 0.35$ exclude the h-LMA region.
Furthermore, precise measurements of CC/NC should strengthen the upper
bounds on mixing and $\Delta m^2$ \cite{ped02b}.

Confronting the prediction for CC/NC with the experimental result
(\ref{cc/nc}) one can understood consequences of (\ref{cc/nc})
immediately.  According to fig.~10 of ref.~\cite{ped02b} (where we
show the contours of constant CC/NC ratio in the $\Delta m^2 - \tan^2
\theta$ plot) the ratio CC/NC decreases with $\Delta m^2$ and $\tan^2
\theta$.  Therefore, with the SNO salt results the best fit point and
the allowed region shift toward the smaller values of $\Delta m^2$ and
$\tan^2 \theta$.  Maximal mixing and the h-LMA region should be
further disfavored.  This is indeed confirmed by the detailed studies
\cite{salt}.

In this paper we present our analysis of the solar neutrino data
including the SNO-II fluxes (sec. 2). We perform a combined analysis
of the solar neutrino and KamLAND results (sec.~3).  Possible effect
of 1-3 mixing is considered in sec.~4.  We show that after the SNO-II
measurements, physics of the solar neutrino conversion is essentially
determined.  In particular, relative contributions of the adiabatic
conversion in matter and the oscillation effect for different energies
can be quantified (sec.~5).  In sec.~6 the expected spectrum
distortion at SNO is studied.  In sec.~7 we give predictions for the
future experiments. The conclusions are presented in sec.~8.

\section{Solar Neutrinos with the SNO salt-phase results}

In the analysis we use the following data set:

\begin{itemize}

\item

3 total rates: (i) the $Ar$-production rate, $Q_{Ar}$, from
Homestake~\cite{Cl}, (ii) the $Ge-$production rate, $Q_{Ge}$, from
SAGE
\cite{sage} and (iii) the combined $Ge-$production rate from GALLEX~\cite{gallex} 
and GNO \cite{gno};

\item

44 data points from the zenith-spectra measured by Super-Kamiokande
during 1496 days of operation \cite{SK,SK2};

\item

34 day-night spectral points from SNO \cite{sno1};

\item

3 fluxes from the SNO salt phase \cite{salt} measured by the CC-, NC
and ES- reactions.  We treat correlations of these fluxes following
prescription in
\cite{snohow}.

\end{itemize}

Altogether the solar neutrino experiments provide us with 84 data
points.

All the solar neutrino fluxes, but the boron neutrino flux, are taken
according to SSM BP2000 \cite{ssm}.  The boron neutrino flux measured
in the units of the Standard Solar Model flux~\cite{ssm}, $f_B \equiv
F_B/F_B^{SSM}$, is treated as a free parameter (here $F_B^{SSM} = 5.05
\cdot 10^{6}$ cm$^{-2}$ s$^{-1}$).  For the $hep-$neutrino flux we
take fixed value $F_{hep} = 9.3 \times 10^{3}$ cm$^{-2}$ s$^{-1}$
\cite{ssm,hepfl} .

We use the same procedure of analysis as in our previous
publication~\cite{us:msw,ped02b}.  In analysis of the solar neutrino
data as well as in the combined analysis of the solar and KamLAND
results we have three fit parameters: $\Delta m^2$, $\tan^2\theta$ and
$f_B$.

We perform the $\chi^2$ test defining
\be
\chi^2_{sun} = \chi^2_{rate} +   \chi^2_{SK} + \chi^2_{SNO-I} + \chi^2_{SNO-II}, 
\label{chi-def}
\ee
where $\chi^2_{rate}$, $\chi^2_{SK}$, $\chi^2_{SNO-I}$ and
$\chi^2_{SNO-II}$ are the contributions from the total rates, the
Super-Kamiokande zenith spectra, the SNO day and night spectra and the
SNO fluxes from the salt phase correspondingly.  The number of degrees
of freedom is 84 - 3 = 81.
 
The minimum, $\chi^2_{sun} (min)/ d.o.f. = 67/81$ is achieved at
\be
\Delta m^2 = 6.31 \cdot 10^{-5} {\rm eV}^2 ,
~~~ \tan^2 \theta = 0.39,~~~ f_B = 1.063.
\label{bfsol}
\ee
From the pre-salt analysis we had: $(\Delta m^2, \tan^2 \theta) =
(6.15 \cdot 10^{-5}, 0.41)$~\cite{us:msw}.  So, the SNO-II data only
slightly shifted the best fit point to smaller values of $\Delta m^2$
and $\tan^2 \theta$.  The boron neutrino flux was smaller: $f_B =
1.05$.

\begin{figure}[ht]
\centering\leavevmode
\epsfxsize=.8\hsize
\epsfbox{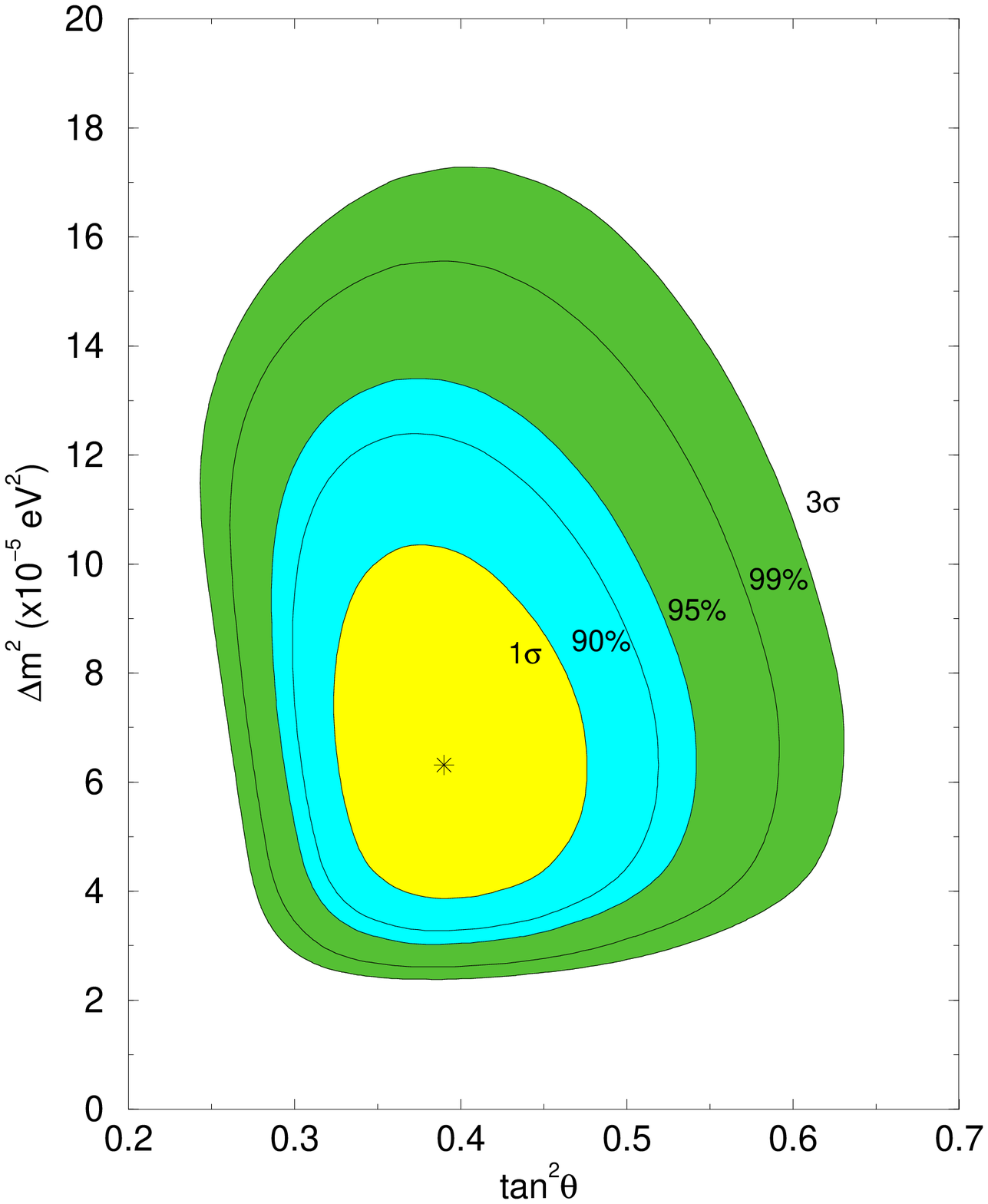}
\caption{The allowed regions in $\tan^2\theta - \Delta m^2$ 
plane, from a combined analysis of the solar neutrino data at
1$\sigma$, 90\%, 95\%, 99\% and 3$\sigma$ C.L..  The boron neutrino
flux is treated as free parameter.  The best fit point is marked by
star.  }
\label{2nusol}
\end{figure}

We construct the contours of constant confidence level in the $\Delta
m^2 - \tan^2 \theta$ plot (fig. \ref{2nusol}) using the following
procedure.  We perform minimization of $\chi^2$ with respect to $f_B$
for each point of the oscillation plane, thus getting
$\chi^2_{sun}(\Delta m^2,
\tan^2 \theta)$.  Then the contours are defined by the condition
$\chi^2(\Delta m^2, \tan^2 \theta) = \chi^2 (min) +
\Delta \chi^2$, where $\Delta \chi^2 = 2.3,\,4.61,\,5.99,\,9.21$ and
$11.83$ are taken for $1\sigma,\,90\%,\,95\%$ and $99\%$ C.L.  and
$3\sigma$.

In contrast to the best fit point, the influence of the SNO-II data on
the size of allowed region, especially for high confidence levels is
much stronger.  The data reduce substantially the allowed region from
the parts of large $\Delta m^2$ and $\tan^2 \theta$.  The borders of
$3\sigma$ region are shifted as $(4.5 \rightarrow 1.6)\cdot 10^{-4}$
eV$^2$ for $\Delta m^2$, and $(0.84 \rightarrow 0.64)$ for $\tan^2
\theta$.

Shift of the $1\sigma$ contour is much weaker.  Projecting the
$1\sigma$ region from fig.~\ref{2nusol} we find the intervals:
\be
\Delta m^2 = (3.8 - 10) \cdot 10^{-5} {\rm eV}^2 ,
~~~ \tan^2 \theta = 0.325 - 0.475.
\label{sol1s}
\ee
The lower bounds for $\Delta m^2$ and $\tan^2 \theta$ are practically
unchanged.\\

\section{Solar neutrinos and KamLAND
\label{solakam}}

The KamLAND data have been analyzed using the Poisson
statistics. Minimizing
\be
\chi^2_{KL} \equiv \sum_{i=1,13} 2 \left[{N_i^{th}-N_i^{obs} + 
N_i^{obs}ln\left(\frac{N_i^{obs}}{N_i^{th}}\right)}\right]
\label{kl}
\ee
we find the best fit point
\be
\Delta m^2 = 7.24 \cdot 10^{-5} {\rm eV}^2,   ~~~~\tan^2\theta = 0.52. 
\label{Kam-par}
\ee

Confronting the oscillation parameters determined independently from
the solar neutrino experiments (\ref{bfsol}) and KamLAND
(\ref{Kam-par}) one can check CPT and search for new physics ``beyond
LMA".  Since there is no significant change in the best fit point and
$1\sigma$ region for solar neutrinos, status of the CPT check is
practically unchanged in comparison with the pre-salt analysis.  The
data are well consistent with the CPT conservation.  As before, there
is an overlap of the $1\sigma$ regions of oscillation parameters found
from the solar data and KamLAND results. The b.f. point (\ref{bfsol})
is at the border of 95\% C.L. region allowed by KamLAND \cite{kamf}.

Assuming CPT conservation we have performed a combined fit of the
solar neutrino data and KamLAND spectral results.  We calculate the
global $\chi^2$
\be
\chi^2_{global} \equiv \chi^2_{sun} +   \chi^2_{KL} ~.  
\label{chi-gl}
\ee
There are 84 (solar) + 13 (KamLAND) data points - 3 free parameters =
94 d.o.f..  The absolute minimum, $\chi^2_{global}(min)/d.o.f. =
73.4/94$ is at
\be
\Delta m^2 = 7.13 \cdot 10^{-5}  {\rm eV}^2,~~~\tan^2 \theta = 0.40,
~~~f_B = 1.038.
\label{bfglob}
\ee 
The SNO salt results have shifted the best global fit point to smaller
$\Delta m^2$ and $\tan^2 \theta$ in comparison with the results of
previous analysis \cite{ped02b}: $\Delta m^2 = 7.30 \cdot 10^{-5} {\rm
eV}^2$, $\tan^2 \theta = 0.41$, $f_B = 1.05$.

\begin{figure}[ht]
\centering\leavevmode
\epsfxsize=.8\hsize
\epsfbox{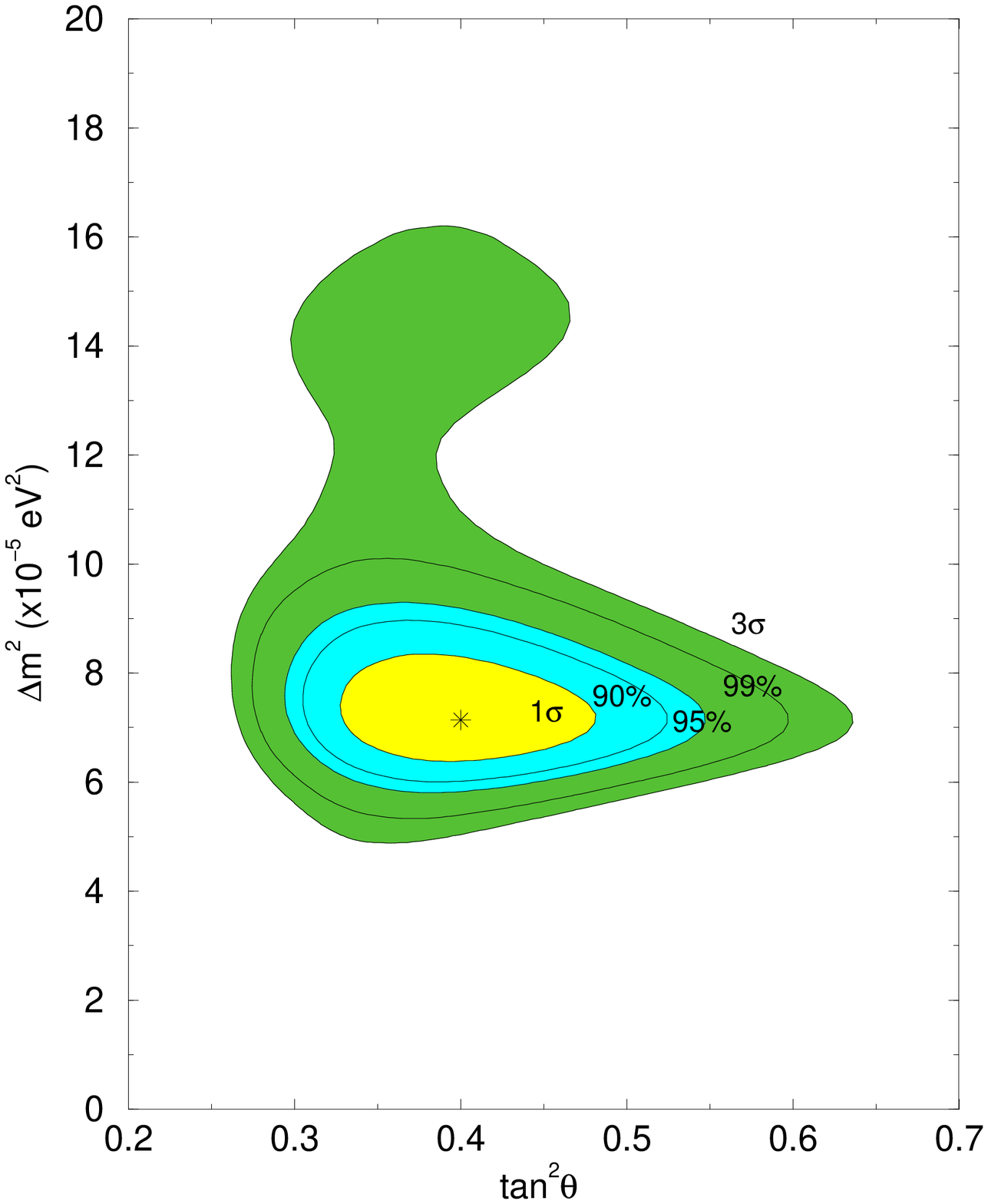}
\caption{The allowed regions in $\tan^2\theta - \Delta m^2$
plane, from a combined analysis of the solar neutrino data and the
KamLAND spectrum at 1$\sigma$, 90\%, 95\%, 99\% and 3$\sigma$ C.L..
The boron neutrino flux is treated as free parameter.  The best fit
point is marked by star.  }
\label{2nusolK}
\end{figure}

We construct the contours of constant confidence level in the $\Delta
m^2 - \tan^2 \theta$ plot (fig. \ref{2nusolK}) using the same
procedure as in sec.~2.  According to fig.~\ref{2nusolK} the intervals
of parameters obtained by projection of the $1\sigma$ allowed region
equal
\be
\Delta m^2 = (6.4 - 8.4) \cdot 10^{-5}  {\rm eV}^2 , 
~~~ \tan^2 \theta = 0.33 - 0.48 .
\label{lower}
\ee  

Notice that inclusion of the KamLAND result does not change the
$3\sigma$ upper bounds on $\Delta m^2$ and $\tan^2\theta$ (compare
with fig.~\ref{2nusol}).  However, KamLAND further strengthen the
lower bound on $\Delta m^2$:
\be
\Delta m^2 > 4.8 \cdot 10^{-5}~~ {\rm eV}^2, ~~~  (3\sigma).   
\label{bfp2}
\ee

As a result of the SNO-II measurements, the h-LMA region, $\Delta m^2
(12 - 16) \cdot 10^{-5}~~ {\rm eV}^2$, is strongly disfavored being
accepted with respect to the global minimum (\ref{bfglob}) at about
$3\sigma$ level only.

From fig.~\ref{2nusolK} we get the following upper bounds on mixing:
\be
\tan^2 \theta < \left\{ 
\begin{array}{ll}
0.48, &~~ 1\sigma \\ 0.55, &~~ 2\sigma \\ 0.64, & ~~ 3\sigma\\
\end{array}
\right. .
\label{mixup}
\ee
Maximal mixing is excluded at $ 5.2\sigma$.  These bounds follow from
the solar neutrino data being practically unaffected by KamLAND.

\begin{figure}[ht]
\centering\leavevmode
\epsfxsize=.8\hsize
\epsfbox{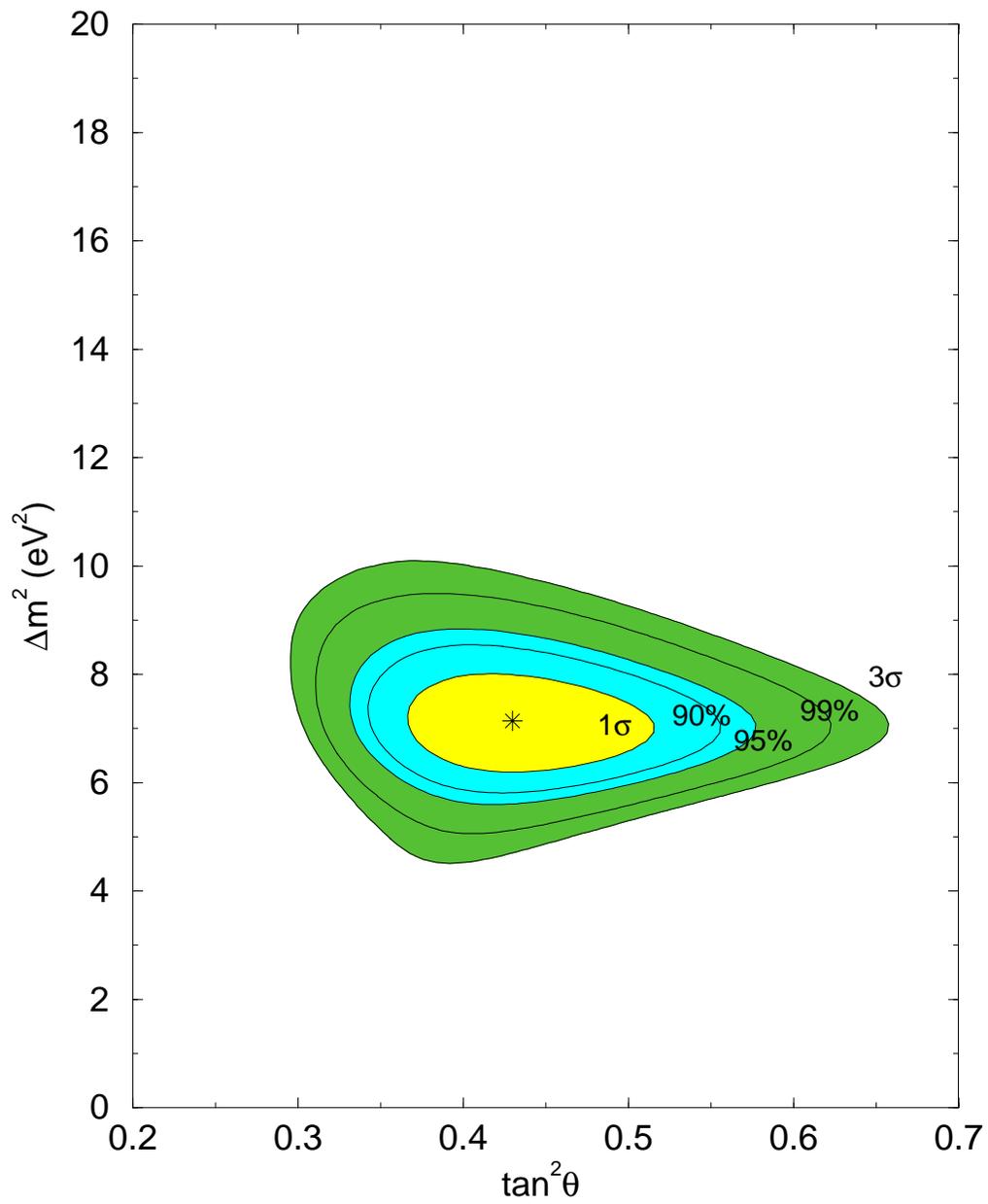}
\caption{
The same as in fig.~\ref{2nusolK} with the SSM predicted value of the
boron neutrino flux.  }
\label{ssmfb}
\end{figure}

To check stability of our results we have performed analysis taking
the boron neutrino flux as it is predicted by the SSM: $f_B = 1$, with
the corresponding theoretical errors~\cite{ssm}.  Results of the
analysis with two free parameters: $\Delta m^2$ and $\tan^2\theta$ are
shown in fig.~\ref{ssmfb}.  In the best fit point:
\be
\Delta m^2 = 7.36 \cdot 10^{-5}  {\rm eV}^2,~~~\tan^2 \theta = 0.41 
\label{newfit}
\ee
we have $\chi^2/d.o.f. = 74.0/82$.  The allowed region is shifted to
larger $\tan^2\theta$ in comparison with the free $f_B$ fit. Indeed,
in the latter case we had $f_B > 1$, so to compensate the decrease of
$f_B$ the survival probability, which is proportional to $\sin^2
\theta$, should increase.  Also the contours of constant confidence
level shift to larger mixings.  The $3\sigma$ upper bound becomes
weaker: $\tan^2 \theta < 0.628$.

Notice that the h-LMA region is rejected now at the $3\sigma$ level.

\section{Effect of 1-3 mixing}

As it was found in the pre-salt phase analysis, the 1-3 mixing
improves the relative goodness of the fit in the h-LMA region (see,
{\it e.g.}, \cite{ped02b}).  Can non-zero 1-3 mixing substantially
change the results of sect. 3?

Both for the KamLAND and for solar neutrinos the oscillations driven
by $\Delta m^2_{13}$ are averaged out and signals are determined by
the survival probabilities
\be
P_{ee} = (1 - \sin^2 \theta_{13})^2 {P}_{2} + \sin^4 \theta_{13}
\approx (1 - 2\sin^2 \theta_{13}) {P}_{2}, 
\label{threepr}
\ee
where $P_{2} = P_{2}(\Delta m^2_{12}, \theta_{12})$ is the two
neutrino vacuum oscillation probability in the case of KamLAND and it
is the matter conversion probability in the case of solar neutrinos.

The effect of non-zero 1- 3 mixing is reduced to the overall
suppression of the survival probability. According to recent analysis
of the atmospheric neutrino data~\cite{atm} the allowed region of
oscillation parameters has shifted to smaller $\Delta m^2_{13}$, and
consequently, the upper bound on 1-3 mixing from the CHOOZ
experiment~\cite{chooz} becomes weaker: $\sin^2 \theta_{13} = 0.067$
($3\sigma$) \cite{atmfo}.  For this value of $\sin^2 \theta_{13}$, the
suppression can reach $13 \%$ in (\ref{threepr}).  The influence of
$\sin^2 \theta_{13}$ on the global fit can be traced in the following
way.

There are three sets of observables for which effects of $\sin^2
\theta_{13}$ are different.

1) Total rates (fluxes) at high energies measured by SNO and SK. These
rates depend essentially on the combination $\cos^4 \theta_{13} P_2$.
In particular,
\be
\frac{\rm CC}{\rm NC} \approx \cos^4 \theta_{13} P_2, ~~~~~~ 
\frac{\rm ES}{\rm NC} \approx \cos^4 \theta_{13} P_2 (1 - r) + r. 
\ee
The ratios of fluxes are unchanged if
\be
\cos^4 \theta_{13} \langle P_2 (\Delta m^2_{12}, \tan^2 \theta_{12} )\rangle  = const.,
\label{invar}
\ee
where $\langle ... \rangle$ is the averaging over the relevant energy
range.  The product (\ref{invar}) is invariant with increase of
$\sin^2 \theta_{13}$ (decrease of $\cos^4 \theta_{13}$) if the
probability increases. In the region near the b.f. point
(\ref{bfglob}) the latter requires increase of $\Delta m^2_{12}$
or/and $\tan^2\theta_{12}$.  Then the absolute values of fluxes can be
reproduced by tunning $f_B$ (in the free $f_B$ fit).

With increase of $\sin^2 \theta_{13}$ the predicted spectrum becomes
flatter which does not change the quality of fit significantly.

The day-night asymmetry decreases with increase of $\Delta m^2_{12}$
which is slightly disfavored by the data. Future more precise
measurements of asymmetry will have stronger impact.

2) Low energy observables, sensitive to $pp$- and $Be$- neutrino
fluxes.  They depend on the averaged vacuum oscillation
probability. In particular, the Ge-production rate is proportional to
\be
Q_{Ge} \propto \cos^4 \theta_{13} (1 - 0.5 \sin^2 2\theta).
\label{germanium}
\ee
Notice that practically there is no dependence of these observables on
$\Delta m^2_{12}$.  Since in the best fit point for zero 1-3 mixing
$Q_{theor} \approx Q_{exp}$, a significant increase of $\sin^2
\theta_{13}$ will lead to worsening of the fit (though small non-zero
1-3 mixing, $\sin^2 \theta_{13} < 0.04$, could be welcomed).

3) KamLAND. The present best fit point is near the maximum of
oscillation survival probability averaged over the atomic reactors.
So, the increase of 1-3 mixing could be compensated by the decrease of
1-2 mixing.

These features allow one to understand results of the data fit.

\begin{figure}[ht]
\centering\leavevmode
\epsfxsize=.8\hsize
\epsfbox{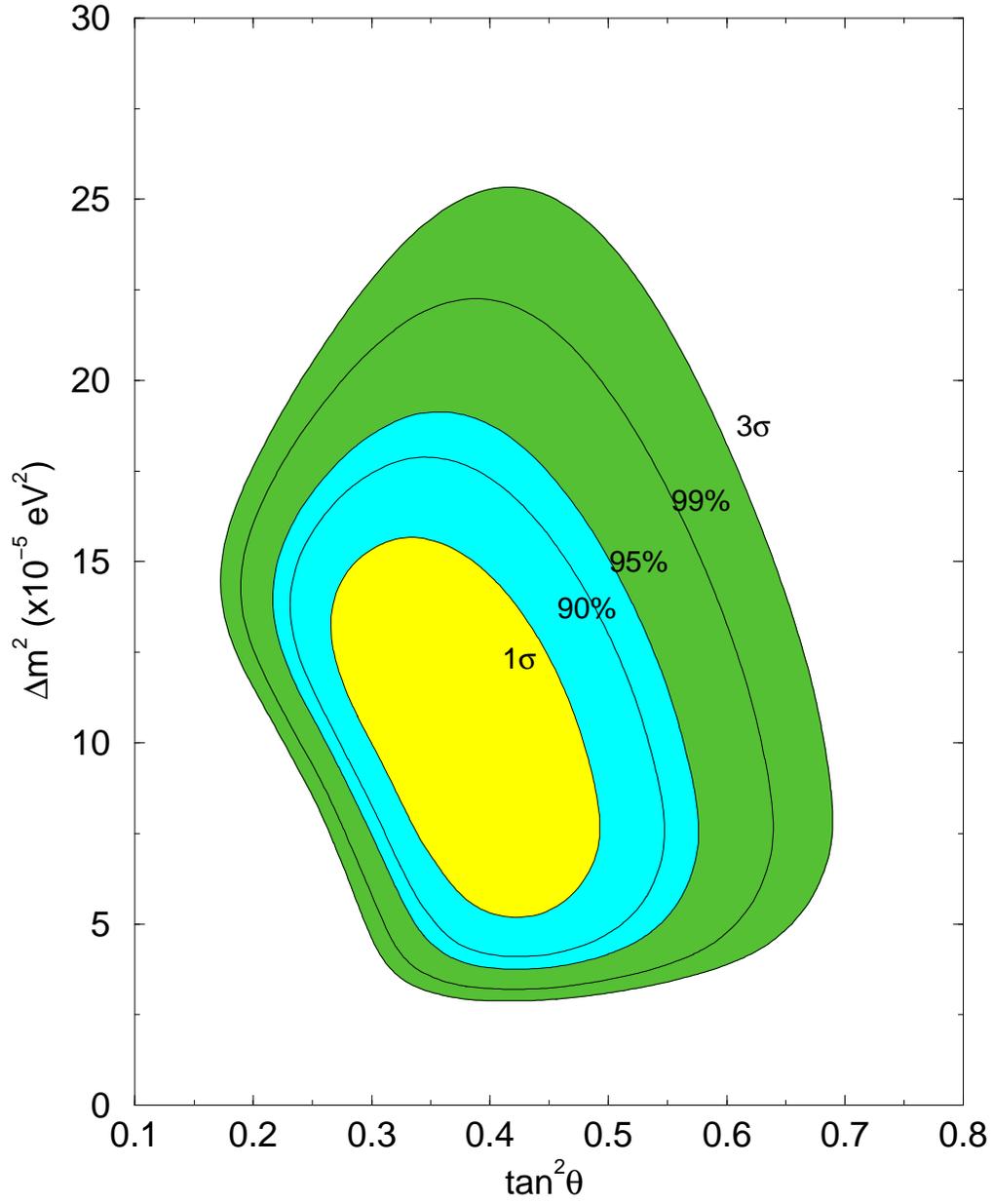}
\caption{Three neutrino analysis with 
$\sin^2\theta_{13}=0.067$.  The allowed regions in $\tan^2\theta -
\Delta m^2$ from a combined fit of the solar neutrino data at the
1$\sigma$, 90\%, 95\%, 99\% and 3$\sigma$ C.L..  }
\label{3nusol}
\end{figure}

We have performed analysis of the solar neutrino data for fixed value
$\sin^2 \theta_{13} = 0.067$ (fig. \ref{3nusol}).  The number of
degrees of freedom is the same as in the $2\nu$ fit and we follow the
procedure described in sect. \ref{solakam} with survival probabilities
given in (\ref{threepr}). The best fit point
\be
\Delta m^2 = 11.0 \cdot 10^{-5}  {\rm eV}^2,~~~\tan^2 \theta = 0.38,
~~~f_B = 1.03
\ee
corresponds to $\chi^2/d.o.f. = 66.8/ 81$. It is shifted to larger
$\Delta m^2$ to satisfy condition (\ref{invar}). The shift to larger
1-2 mixing is disfavored by the Gallium experimental results
(\ref{germanium}). The allowed region is also shifted to larger
$\Delta m^2$ and its size increases (also in the mixing
direction). These results agree with the analysis in
\cite{valle2}. 

We also performed the combined analysis of the solar neutrino data and
the KamLAND spectrum for $\sin^2 \theta_{13} = 0.067$.  In the best
fit point we find $\chi^2_{min}/d.o.f. = 74.3/94$ and
\be
\Delta m^2 = 7.1  \cdot 10^{-5}  {\rm eV}^2 , 
~~~ \tan^2 \theta = 0.42, ~~f_B = 1.03.
\label{3spel}
\ee  
So, introduction of the 1-3 mixing slightly worsen the fit: $\Delta
\chi^2 \approx 1.9$. The best fit point shifts to larger $\tan^2
\theta$ 
to satisfy the condition
(\ref{invar}). Significant shift to larger $\Delta m^2$ is not allowed
by the KamLAND spectral data.  At the same time, the 1-3 mixing
improves a fit in the h-region.  This region is accepted now at 90 \%
C.L. with respect to the best fit point (\ref{3spel}).

Notice however, that such an improvement is realized for values of
$\sin^2\theta_{13}$ which are accepted by the CHOOZ and atmospheric
data at the $3\sigma$ level.  So, inclusion of the CHOOZ result in the
global fit changes a situation and no significant improvement of the
fit in the h-LMA region occurs \cite{choubey}. In fact, at $\sin^2
\theta_{13} = 0.067$ this region disappears even at the $3\sigma$
level.

\section{Physics of the solar neutrino conversion.}  

\subsection{Dynamics of conversion}

With new SNO results the h-LMA region is excluded or strongly
disfavored and also significant deviation of 1-2 mixing from maximum
is established.  This essentially determines both qualitatively and
now quantitatively the physical picture of solar neutrino conversion
(see also \cite{venice}).  Different analyzes of data converge to
\be
\Delta m^2 = (6 - 8) \cdot 10^{-5}  {\rm eV}^2 ,
~~~ \sin^2 \theta = 0.28 - 0.30.
\label{conv}
\ee
The non-zero 1-3 mixing may produce some small shift of the
parameters.  Since the physical picture is nearly the same for all
points in these intervals, for further estimations we will take the
values (\ref{bfglob}) as the reference point.

We take the difference of the matter potentials for $\nu_e$ and
 $\nu_a$ (active) according to the Standard Model:
\be
V = \sqrt{2} G_F \rho Y_e/m_N.
\label{poten}
\ee
Here $\rho$ is the matter density, $Y_e$ is the number of electrons
per nucleon, and $m_N$ is the nucleon mass.  In fact, the latest
experimental data allow to check the presence of such a potential in
the model independent way. The extracted value of the potential is in
agreement with (\ref{poten}) (within $1\sigma$) and $V = 0$ is
rejected at $\sim 5.6 \sigma$ level~\cite{fogli}.

For parameters (\ref{conv}) the neutrino evolution inside the Sun
occurs in the highly adiabatic regime.  It is convenient to write the
averaged adiabatic survival probability as~\cite{w,ms}
\be
P = \sin^2\theta + \cos 2\theta ~\cos^2\theta_m^0,
\label{adiab}
\ee
where $\theta_m^0$ is the mixing angle in the production
point. Provided that there is no coherence lost (see discussion at the
end of the section), the
depth of oscillations at the surface of the Sun equals~\cite{ms}
\be
A_P = \sin 2\theta \sin 2\theta_m^0.
\ee
So that the probability (being the oscillatory function of distance)
is inscribed in the following oscillation strip
\be
P \pm \frac{1}{2} \sin 2\theta \sin 2\theta_m^0.
\ee

The oscillation length in matter, $l_m$, is smaller than the resonance
oscillation length, $l_R \equiv l_{\nu}/\sin 2\theta$, where $l_{\nu}$
is the vacuum oscillation length.  So, for typical energy 10 MeV we
find $l_m \sim 3.5 \cdot 10^{7}$ cm. Along the solar radius,
$R_{\odot}$, about $R_{\odot}/l_m \sim 2 \cdot 10^{3}$ oscillation
length are obtained, and consequently, the oscillations are strongly
averaged out.

Dynamics of the effect depends on $\sin^2 \theta$ and
\be
\eta (E) \equiv \frac{E_{kin}}{V_0} = \frac{\Delta m^2}{2EV_0} =   
\frac{\Delta m^2 m_N}{2\sqrt{2} E G_F (\rho Y_e)_0}, 
\label{eta}
\ee
which is the ratio of the ``kinetic'' energy, $\Delta m^2/2E$, and the
potential energy in the neutrino production point, $V_0$.  For a given
energy, $\eta$ determines the relative contributions of the vacuum
oscillation and the matter effect.

\subsection{Two limits}

Depending on $\eta$ there are two limits:

{\it 1. Matter dominance}: $\eta \ll 1$ which corresponds to $E
\rightarrow \infty$.  In this limit the neutrino flavor evolution has
a character of the non-oscillatory ($A_P = 0$) conversion with the
survival probability ~\cite{ms}:
\be
P_{non-osc} = \sin^2 \theta .
\label{p-non}
\ee 
In this case neutrinos produced far above the resonance density
propagate to zero (small) density adiabatically.  The initial mixing
is strongly suppressed ($\theta_m^0 \approx \pi/2$) and the
propagating neutrino state practically coincides with the heaviest
eigenstate: $\nu(t) = \nu_{2m}$. At the exit from the Sun: $\nu(t) =
\nu_{2}$ \cite{ms}.

{\it 2. Vacuum dominance:} $\eta \gg 1$ which corresponds to $E
\rightarrow 0$.  Matter effects are small. The flavor evolution has a
character of vacuum oscillations, so $P$ converges to the averaged
oscillation probability
\be
P \rightarrow P_{vac} = 1 - 0.5 \sin^2 2\theta,
\label{vac-p}
\ee
and $A_P \rightarrow \sin^2 2\theta$.

For the reference value of mixing (\ref{bfglob}) we find
\be
P_{non-osc} = 0.281, ~~~~ P_{vac} = 0.596.
\label{evalua} 
\ee

The resonance value of $\eta$, for which $P = 1/2$, equals
\be
\eta_R = \frac{1}{\cos 2\theta} = 2.28. 
\label{reseta}
\ee
It marks the transition region between the two extreme cases.

The boron neutrinos are produced in the region where the density $\rho
Y_e$ decreases from 100 to 70 g/cc. The layer with maximum of
emissivity ($r = 0.043 R_{\odot}$) has the effective density $\rho Y_e
\sim 93.4$ g/cc \cite{ssm}. For this density the value (\ref{reseta})
corresponds to the resonance energy $E_R = 2.2$ MeV.  Neutrinos with
$\eta <
\eta_R$ ($E > 2.2$ MeV) are produced  above the resonance (density) and
then propagating to the surface of the Sun cross the resonance layer.
Neutrinos with $\eta > \eta_R$ ($E < 2.2$ MeV) are produced below the
resonance (density) and never cross the resonance (or cross the
resonance layer twice depending on direction of propagation).

We can also define the ``median'' energy at which the probability
equals the average of the two asymptotics (\ref{p-non}) and
(\ref{vac-p}): $P_{med} = 0.5(P_{vac} + P_{non-osc})$. According to
(\ref{adiab}), this corresponds to $\cos^2 \theta_m = 0.5 \cos^2
\theta$ and is realized for
\be
\eta_{med} = \left[ 
\cos 2\theta + \frac{2 \sin^3\theta}{\sqrt{1 + \sin^2 \theta}}\right]^{-1}.
\ee
We find that $\eta_{med} = 1.43$ and $E_{med} = 3.54$ MeV.

The ratio CC/NC measured by SNO determines the survival probability
averaged over the energy range above the SNO threshold for the CC
events ($E \sim 5$ MeV):
\be
\frac{\rm CC}{\rm NC} = \langle P \rangle . 
\ee
For the reference value (\ref{bfglob}) we get $\langle P \rangle =
0.322$. This value is slightly larger than the experimental result
(\ref{cc/nc}).

So, both the experimental and theoretical values of CC/NC are rather
close to $P_{non-osc}$ which means that at high energies ($E > 5$ MeV)
the evolution of neutrino state is nearly non-oscillatory conversion.
The difference
\be
\frac{\rm CC/NC}{P_{non-osc}} - 1 = \frac{\rm CC/NC}{\sin^2 \theta}-1 = 0.13  
\ee
is due to

\begin{itemize} 

\item
the averaged oscillation effect inside the Sun. In fact, for $(\rho
Y_e)_0 \approx 93.4$ g/cc and energies relevant for the SNO CC signal,
$E = 14, 10, 6$ MeV, we obtain $\eta = 0.36, 0.50, 0.84$
correspondingly.  These values of $\eta$ are not small, though being
smaller than the resonance value.  Therefore one may expect a
significant deviation of $P$ from the asymptotic value.

\item

the earth regeneration effect.

\end{itemize}

\subsection{Survived probability: adiabatic conversion versus oscillations}

The survival probability can be written as
\be
P = \sin^2\theta + \Delta P_{reg} + \Delta P_{osc},
\ee
where $\Delta P_{reg}$ is the regeneration correction and $\Delta
P_{osc}$ is the oscillation correction.  Both these corrections are
positive.

The regeneration effect, $\Delta P_{reg}$, can be expressed in terms
of the Day-Night asymmetry, $A_{DN}$, as:
\be
\Delta P_{reg} \approx A_{ND} \cdot P_{non-osc} \approx \sin^2\theta \cdot A_{ND} .  
\ee
For the best fit point the asymmetry equals $A_{ND} = 3.0 \%$, and
consequently, $\Delta P_{reg} = 0.008$.

Using formula for the adiabatic conversion (\ref{adiab}) we find:
\be
\Delta P_{osc} = \cos 2\theta \cos^2\theta_m^0 .  
\label{osc-cor}
\ee
Notice that $\cos^2\theta_m^0$ gives the probability to find the
eigenstate $\nu_{1m}$ in the adiabatically propagating neutrino state:
$|\langle \nu_{1m} | \nu (t)\rangle|^2 = \cos^2\theta_m^0 = const$.
In the non-oscillatory limit we would have $\cos^2\theta_m^0 \approx
0$ and $\nu (t) = \nu_{2m}$.  The presence of second eigenstate,
$\nu_{1m}$, in $\nu (t)$, leads to the interference effect, and
consequently, to oscillations.

In the limit of small $\eta$ we find
\be
\cos^2\theta_m^0 \approx \frac{1}{4} \sin^2 2\theta~ \eta^2 , 
\ee
and therefore the correction to probability can be written as
\be
\Delta P_{osc}  = \frac{1}{4} \cos 2\theta \sin^2 2\theta ~(\eta^2 + O(\eta^3)). 
\label{corr-t}
\ee
Notice that the correction is quadratic in $\eta$, and furthermore, it
contains small pre-factor $\cos 2\theta/4 \sim 0.1$. It is for this
reason the correction is rather small in spite of large values of
$\eta$. However, convergence of the series is determined by $\eta$
itself, and so, the corrections to the first order $\Delta P_{osc}$
are not small.  Although (\ref{corr-t}) allows to understand the size
of the correction, in our estimations we use exact expression for
$\cos^2\theta_m^0$.  In the limit of small $\eta$ the depth of
oscillations
\be
A_{P} \approx \sin^2 2\theta ~\eta
\ee
decreases linearly, that is, slower than correction to the average
value.

\begin{figure}[ht]
\centering\leavevmode
\epsfxsize=.8\hsize
\epsfbox{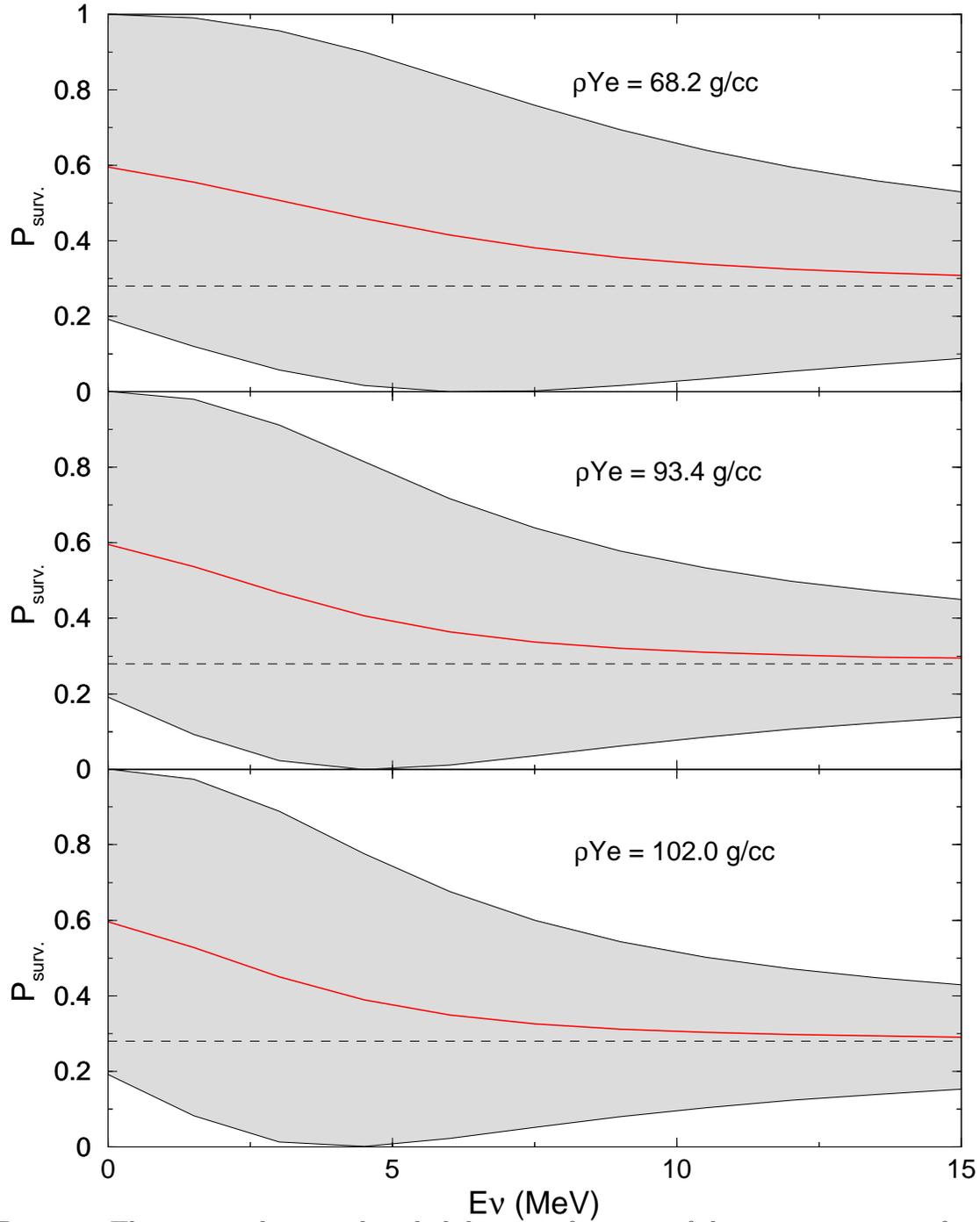}
\caption{The averaged survival probability  as  a function of
the neutrino energy for different production initial densities $\rho
Y_e$.  The oscillation strips (shadowed) show the depth of
oscillations at the surface of the Sun.  The non-oscillatory
conversion probability $P_{non-osc} = \sin^2 \theta$ is shown by the
dashed line.}
\label{prob}
\end{figure}

In fig.~\ref{prob} the averaged survival probability is shown as a
function of the neutrino energy for different production points
(different initial densities).  The shadowed strips show the depth which
oscillations would have at the surface of the Sun, provided that there
is no loss of coherence. The average $P$ converges to
$P_{non-osc}$ (dashed line) with increase of energy and $\rho
Y_e$. The decrease of oscillation depth with $\eta$ is much slower
than convergence of $P$ to $\sin^2\theta$: $\Delta P \propto \eta^2$,
$A_{P} \propto \eta$. The depth of oscillations increases with
decrease of E approaching the vacuum value. Notice that even for the
highest energies of the spectrum the conversion is not completely
non-oscillatory, though $P \approx \sin^2\theta$.

Using $\eta = 0.50$ which corresponds to a typical energy of the
spectrum measured by SNO, $E = 10$ MeV, we get $\Delta P_{osc} \approx
0.030$ (approximate formula (\ref{corr-t}) leads to $\Delta P_{osc}
\approx 0.022$).  This together with the regeneration effect
reproduces well the observed difference of $P$ and $P_{non-osc}$.  The
depth of oscillations for this set of parameters is rather large: $A_P
= 0.45$.

Thus, at energies relevant for the SNO CC events, the survival
probability is about 12\% larger than the non-oscillatory probability.
Oscillations give dominating effect in this difference.  The
regeneration contributes about 4\%.

At low energies ($E < 2$ MeV) the Earth regeneration ($\propto (\Delta
m^2/E)^2$) can be neglected and the probability is given by the vacuum
oscillation formula with small matter corrections.  For $\eta \gg 1$
we can write:
\be
P \approx P_{vac} - \frac{1}{2\eta} \cos 2\theta \sin^2 2\theta .
\label{vac}
\ee
For the beryllium neutrinos the effective density in the production
region $(\rho Y_e)_0 = 87$ g/cc, and correspondingly, $\eta =
6.28$. Inserting this value of $\eta$ in (\ref{vac}) we find $\Delta P
= - 0.028$ which is smaller than 5\% of $P_{vac}$.

For the $pp$-neutrinos the effective density in the region of the
highest production rate is $(\rho Y_e)_0 \sim 68$ g/cc. At $E = 0.4$
MeV this gives $\eta = 17.3$ and correction $\Delta P = - 0.01$.

The results for different values of $\Delta m^2$ can be immediately
obtained from the fig.~\ref{prob} by rescaling of the energy: for
$\Delta m^{'2}$ the probability at the energy $E'$ equals $P(E') =
P(E)$ where $E = E' \cdot (\Delta m^2/ \Delta m^{'2})$.  If, {\it
e.g.}, $\Delta m^{'2} = 14 \cdot 10^{-5}$ eV$^2$, we find from
fig.~\ref{prob}, that $P = 0.46,~ 0.40,~ 0.35$ for $E = 6,~10,~ 14$
MeV. Notice that even for so large $\Delta m^2$, the probability is
substantially lower than the vacuum value and for high energies the
matter effect still dominates.


\subsection{Coherence loss}

Let us consider  effect of the coherence loss of a neutrino state 
on the conversion picture.
The loss of coherence  suppresses  depth of oscillations, so that
the probability converges to the average value.

The coherence length, $L_{coh}$, can be estimated from the condition:
\be
\int^{L_{coh}} dx \Delta v_m = \delta_\nu,
\ee
where  $\Delta v_m \equiv v_{2m} - v_{1m}$ is the difference
of the group velocities of the eigenstates, 
$\delta_\nu$ is the length of the wave packet and $x$ is the distance.
The group velocities equal:
\be
v_{im} = \frac{d H_{im}}{dp}, ~~~ i = 1,2,
\ee
where $H_{i}$ are the eigenvalues of the effective Hamiltonian of the
neutrinos in matter.
For the difference of the velocities we find~\cite{prog}:
\be
\Delta v_m = \frac{\Delta m^2}{2 p^2}
\frac{\eta_x - \cos 2\theta}{\sqrt{\eta_x^2 - \eta_x \cos 2\theta +1}}, 
\label{delta}
\ee
where $\eta_x \equiv \Delta m^2/2pV(x)$.

Let us analyze the expression in (\ref{delta}). 
In vacuum, $\eta_x \rightarrow \infty$ and
\be
\Delta v_m = \Delta v = \frac{\Delta m^2}{2 p^2}.
\ee
With increase of density the difference  $\Delta v_m$ decreases. 
In the resonance we find
\be
\Delta v_m = \frac{\Delta m^2}{2 p^2} \sin 2\theta . 
\ee
$\Delta v_m$  changes the sign at $\eta_x = \cos 2\theta$ 
which corresponds to density $\rho_0 = \rho_R/\cos^2 2 \theta$ 
larger than the resonance density. For $\rho > \rho_0$ the difference
of velocities changes sign.
At very high densities, $\eta_x \rightarrow 0$, we have 
\be
\Delta v_m = - \frac{\Delta m^2}{2 p^2} \cos 2\theta. 
\ee
Notice that in the resonance channel always $|\Delta v_m| < |\Delta v|$. 
So, the matter suppresses the divergence of packets. Furthermore, due
to change of sign of
$\Delta v_m$ the overlap of packets can be recovered. 
If neutrino propagates from high densities (above the resonance)
the packets  first diverge, but then below $\rho_0$ the overlap 
and therefore the coherence can be restored again.

For the smallest value of $\eta_V \sim 0.36$ in our case ($E = 14$
MeV, $\rho Y_e = 94$ g/cc) we get
$\Delta v_m = - 0.08\Delta v$. So,  
the divergence effect can be estimated using the vacuum value 
\be
L_{coh} = {\delta_{\nu}}\frac{2 E^2}{\Delta m^2}. 
\label{cohl}
\ee 

The sizes of wave packets are different for different components of the solar 
neutrino spectrum. Let us consider them in order. 

The lifetime of isolated $^8 B$ nuclei, $\tau_B = 10^{-2}$ s, is very
long and the size of the wave packet is determined by the average time
between two consequent collisions,  $t_{coll}$:   
\be
\delta_{\nu} = c t_{coll}. 
\label{cond}
\ee
Furthermore, the collisions with large enough 
momentum transfer $\Delta p$: 
\be
\Delta p \geq \frac{2\pi}{\delta_{\nu}}
\label{deltap}
\ee 
should be considered. 
For smaller $\Delta p$ the collision does not break coherence (shorten
the packet).  Using Eq.  (\ref{cond}) and (\ref{deltap}) we find the
condition for $\delta_{\nu}$:
\be
\delta_{\nu} = c t_{coll} (\Delta p > 2\pi/ \delta_{\nu}).
\label{cond1}
\ee

The collision time can be estimated as 
\be
t_{coll} = \frac{1}{\sigma (\Delta p) v n } ~, 
\label{coll}
\ee
where $v$ is the velocity of colliding particles (protons, electrons,
the helium atoms), $n \sim \rho/m_N$ is the number density of these
particles and $\sigma(\Delta p)$ is the cross-section. The dominating
effect is the Coulomb scattering, so that the cross-section can be written as 
\be
\sigma{(\Delta p > 2\pi/\delta_{\nu})} = \frac{A}{(\Delta p)^2} = 
\frac{A \delta_{\nu}^2}{(2\pi)^2}, 
\label{crsec}
\ee
where in the lower approximation 
$A \simeq \pi \left(Z\alpha/v\right)^2$ corresponds to the Rutherford
formula, where $Z=5$ is the charge
of Boron nuclei, $\alpha \equiv e^2/4\pi$ and $v$ is the velocity of
a colliding particle. 
In our case $Z\alpha/v>1$ and the perturbation theory is violated.
For estimations
we take $A\sim 0.1 - 1$.
Inserting (\ref{crsec}) into (\ref{coll}) and then (\ref{cond1}) we get 
\be 
\delta_{\nu} = \left[ \frac{4 \pi^2}{ n v A} \right]^{1/3}
\label{de}
\ee
which leads to  
\be
\delta_{\nu} = (1-2) \times 10^{-7} {\rm cm} \,.
\ee
This result is in agreement with results obtained by other
methods~\cite{nuss1,loeb,nuss}. 
The coherence length equals: 
\be
L_{coh} \sim (4-8) R_{\odot} \left(\frac{E}{10 {\rm MeV}} \right)^2 .
\ee
Here $R_{\odot} = 7 \cdot 10^{10}$ cm is the solar radius. 
The matter effect leads to suppression of the divergence, and
consequently, to larger coherence length.
According to (\ref{de}) for the observable part of the boron neutrino
spectrum the effect of the coherence loss is small. Strong divergence
of packets in the conversion region ($x \sim 0.1 R_{\odot}$) occurs
for $E < 1.5$ MeV.

For the $^7Be$ neutrinos the length of wave packet is determined by the
production time~\cite{nuss}, that is by the capture of the electron.
The latter is given by the time an electron wave packet of the size 
$\delta_e$ crosses nucleus~\cite{nuss}: 
\be
t = \frac{\delta_e}{v_e}\,\, . 
\label{capt}
\ee 
Using the thermal velocity: $v_e \sim \sqrt{3kT/m_e}$ and the thermal
wave length for the electron packet: $\delta_e \sim 2\pi/m_e v_e$, we
get from (\ref{capt}) for the neutrino wave packet length:
\be
\delta_{\nu} \sim  \frac{2\pi}{m_e v_e^2} =  \frac{2\pi}{3 k T}~. 
\label{pack}
\ee
Numerically this gives $\delta_{\nu} \sim 6\cdot 10^{-8}$ cm \cite{nuss}. 

The $Be$ neutrinos undergo nearly vacuum oscillations with small
matter effect, so to calculate the coherence length we can use the
formula (\ref{cohl}): \be L_{coh} (Be) \sim 0.01 R_{\odot}.  \ee That
is, the coherence is lost already in the production region. Notice
that the oscillation length is even smaller ($3 \cdot 10^{6}$ cm) and
$L_{coh}/l_{\nu} \sim 200$.

For the $pp$-neutrinos the coherence length is given by (\ref{cohl}).
The size of the packets is determined by the time of interaction
which, in turn, is given by the wave packet of the colliding proton,
similarly to the case of $^7 Be$ neutrinos. Notice that expression
(\ref{pack}) does not depend on a mass of colliding particle, and
therefore it can be immediately applied for the $pp$-neutrinos.  As a
result, the neutrino wave packet has the same size $\delta_{\nu} \sim
6\cdot 10^{-8}$ cm.  The coherence length is shorter due to smaller energies: 
\be
L_{coh} (pp) \sim 0.002 R_{\odot} \left(\frac{E}{0.4 {\rm MeV}}\right)^2.
\label{coh2}
\ee
Again the loss of coherence occurs already in the  production region. 
The number of oscillation periods before complete averaging:
$L_{coh}/l_{\nu} < 100$.

\section{Spectrum distortion at SNO}

Distortion of the energy spectrum is the generic consequence of the
LMA MSW solution.  As we discussed in the previous section, with
decrease of energy the survival probability increases due to increase
of the oscillation contribution.

The energy spectrum of electrons has been calculated according to
\be
S(T_{eff})=\int_{T_r}\int_{E} \frac{d\sigma(T_r,E)}{dT_r}
R(T_r,T_{eff})
\left[ P^{B}f_B\phi_{B}(E) + P^{hep}\phi_{hep}(E)\right] 
dE \, dT_r ~~,
\label{spectr}
\ee
where $T_{eff}$ is the measured electron kinetic energy, $T_r$ is the
real electron kinetic energy, $E$ is the neutrino energy,
$\phi_{B(hep)}$ is the Boron (hep) neutrino flux,
$d\sigma(T_r,E)/dT_r$ is the differential cross section taken from
\cite{cross}, and $R(T_r,T_{eff})$ is the SNO resolution function,
$P^{B}$ and $P^{hep}$ are the survival probabilities averaged over the
corresponding production regions.

\begin{figure}[ht]
\centering\leavevmode
\epsfxsize=.8\hsize
\epsfbox{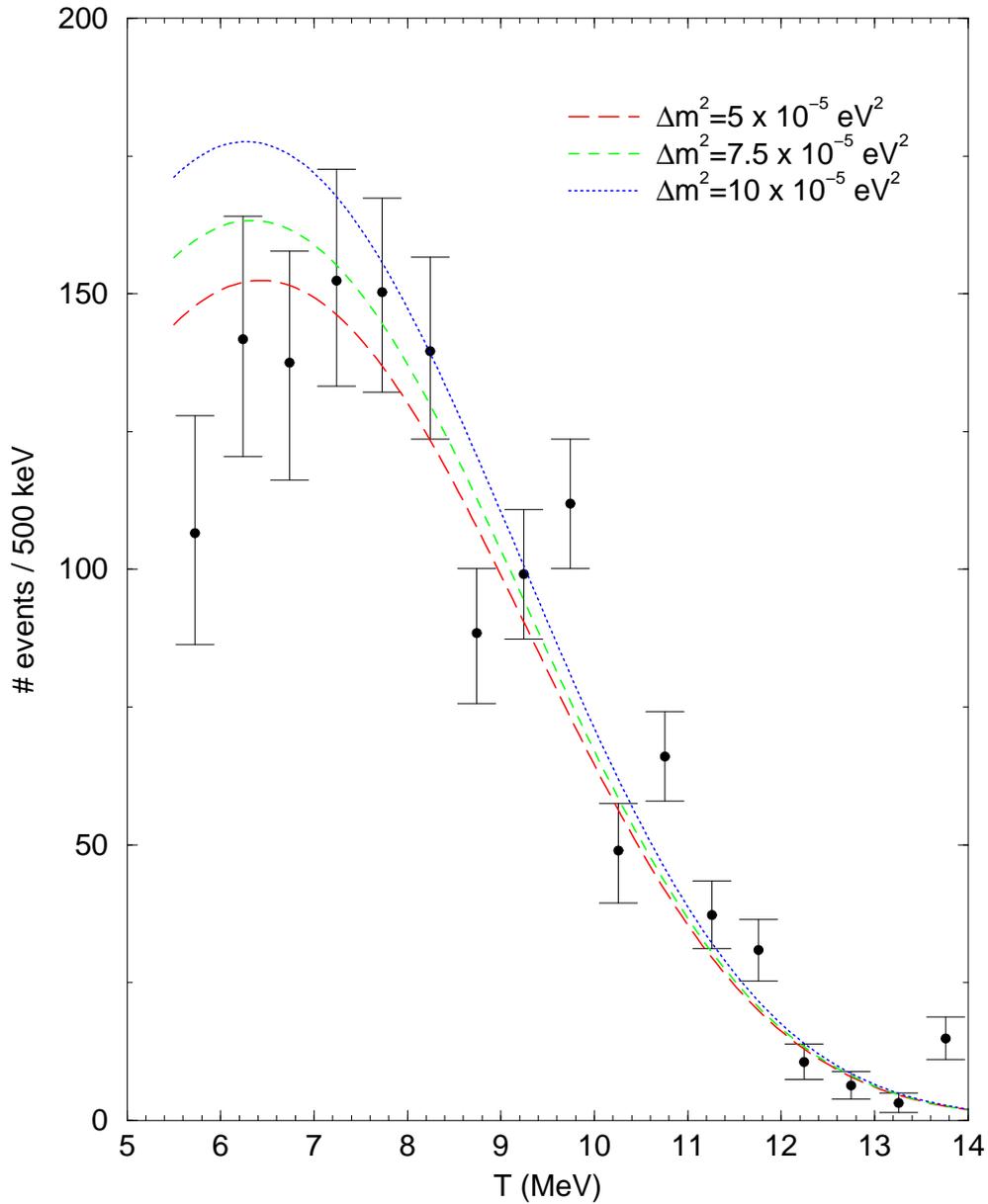}
\caption{The energy spectrum of the CC reaction events. 
Shown is the distribution of events in the kinetic energy for
different values of $\Delta m^2$ and $\tan^2\theta = 0.39$.  We show
also the SNO experimental points from the salt phase.}
\label{spec}
\end{figure}

In fig.~\ref{spec} we show the results of calculations of spectra
for different values of $\Delta m^2$.  The distortion due to
oscillations which dominates at low energies is partly compensated by
the regeneration effect at high energies.  Thus, for the day signal
one expects stronger upturn.

According to fig.~\ref{spec}, the upturn is about 8 - 15\%.  We show
also the SNO experimental points from the salt phase.  A dependence of
the distortion on $\Delta m^2$ is rather weak in the allowed region.
Notice that in the low energy part the spectrum has a tendency to turn
down in contrast to the expected effect.  At $E < 7.5$ MeV the points
are systematically below the predicted rate.  One should notice,
however, that the experimental points include the statistical error
only and it is not excluded that some systematics explains the
observed result.

The same effect - an absence of the upturn of the spectrum - is
observed in the phase -I of the SNO experiment~\cite{sno1}. The
spectral data agree well with the undistorted spectrum.  It would be
interesting to combine the results of both phases to improve
statistics (and probably reduce the number of bins). Being confirmed,
the fact of absence of the upturn or even a turn down at low energies
can be explained by the effect of additional sterile
neutrino~\cite{ster}.

\section{Next step} 

Let us present predictions for the forthcoming experiments.

\begin{figure}[ht]
\centering\leavevmode
\epsfxsize=.8\hsize
\epsfbox{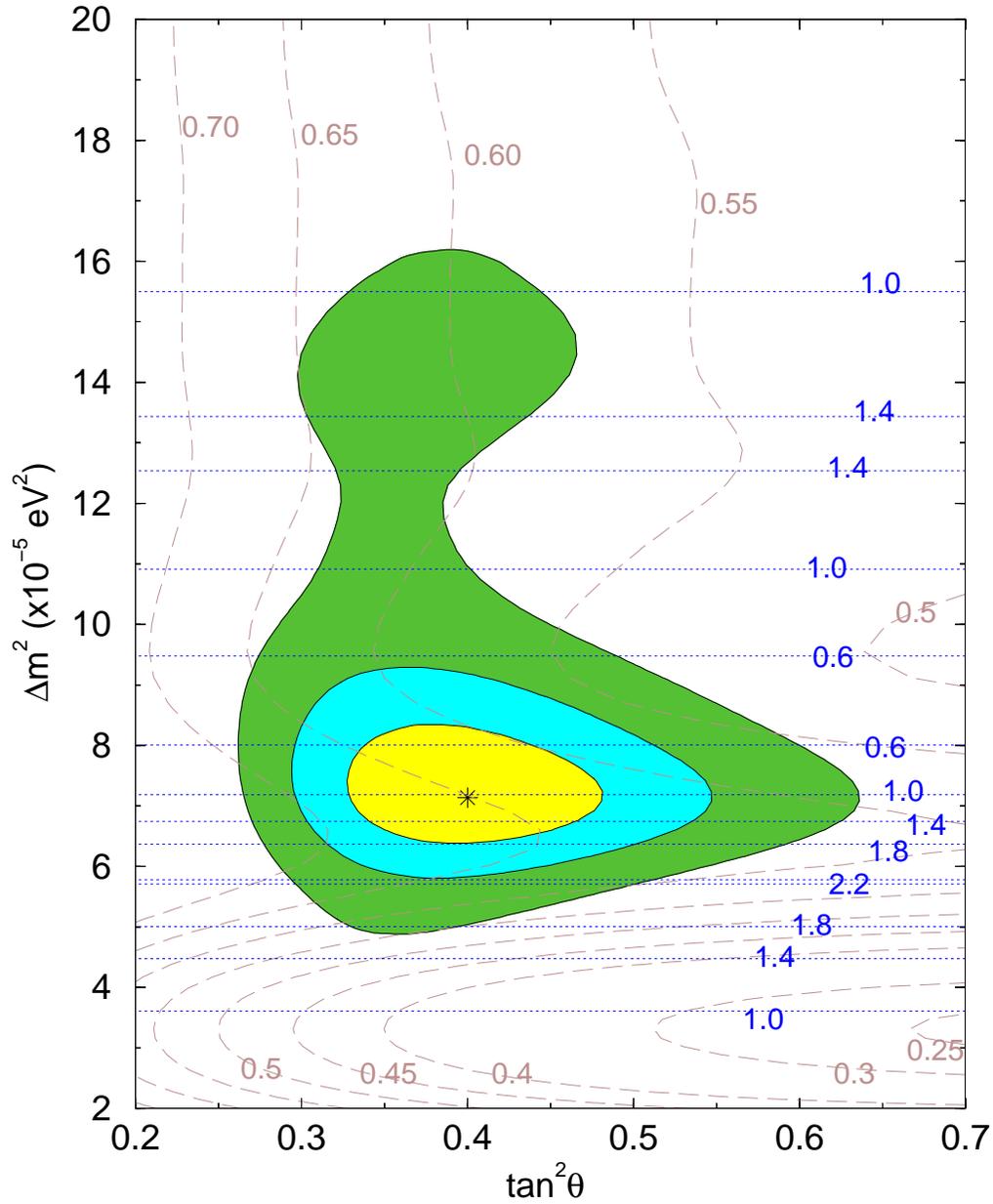}
\caption{The contours of constant rate suppression, $R_{KL}$ (dotted) and the spectrum 
shape parameter $k$ (dashed) at KamLAND.  We show also the allowed
regions of the oscillation parameters from the combined fit of the
solar neutrino data and the KamLAND spectrum. The best fit point is
indicated by a star.}
\label{fig:kamland}
\end{figure}

{\it 1). KamLAND}.  More precise measurements of the rate and spectrum
distortion are expected.  That can further diminish uncertainty in the
determination of $\Delta m^2$.  In fig.~\ref{fig:kamland} we show the
contours of constant suppression of the KamLAND rate $R_{KL}$ above
2.6 MeV:
\be 
R_{KL} \equiv \frac{N(\Delta m^2, \tan^2\theta)}{N_0},
\ee
where $N$ and $N_{0}$ are the numbers of events with and without
oscillations.  As follows from the figure strengthening of the lower
bound on $R_{KL}$ will cut the allowed region from the side of small
$\Delta m^2$ ($ \sim 5 \cdot10^{-5}$ eV$^2$) as well as large $\Delta
m^2$ ($ \sim 9 \cdot10^{-5}$ eV$^2$) and large mixings: $\tan^2 \theta
\sim 0.45$. In contrast, strengthening of the upper bound on $R_{KL}$
will disfavor the region of small mixings: $\tan^2 \theta \sim 0.3$.

The spectrum distortion can be characterized by a relative suppression
of rates at the high and low energies. We choose $E = 4.3$ MeV as the
border line ~\cite{ped02b}, so that the interval (2.6 - 4.3) MeV
contains 4 energy bins.  Introducing the rates $R_{KL} (< 4.3~{\rm
MeV})$ and $R_{KL} (> 4.3~{\rm MeV})$ we define the {\it shape
parameter} as
\be
k = \frac{1 - R_{KL} (> 4.3~{\rm MeV})}{1 - R_{KL} (< 4.3~{\rm MeV})}.
\ee
$k$ does not depend on the normalization of spectrum and on the mixing
angle in the $2\nu$ context.  It increases with the oscillation
suppression of signal at high energies. $k > 1$ ($k < 1$) means
stronger suppression at high (low) energies.  The present KamLAND data
give~\cite{ped02b}
\be 
k^{exp} = 0.84^{+0.42}_{-0.35}, ~~~1\sigma.
\label{k-exp}
\ee
In fig.~\ref{fig:kamland} we show the contours of constant shape
parameter.  According to this figure at the $1\sigma$ level
\be
k^{th} = 1.05^{+0.75}_{-0.50}
\label{k-th}
\ee 
in a very good agreement with (\ref{k-exp}).  In the l-LMA ($1\sigma$)
region we find $k = 0.5 - 2.3$.  So that even mild increase of
statistics will influence the allowed range of $\Delta m^2$.

In the h-LMA region the allowed interval, $k = 0.9 - 1.4$, is
narrower.  So, if forthcoming measurements favor $k < 0.9$ or $k >
1.4$, the h-LMA region will be further discriminated.

\begin{figure}[ht]
\centering\leavevmode
\epsfxsize=.8\hsize
\epsfbox{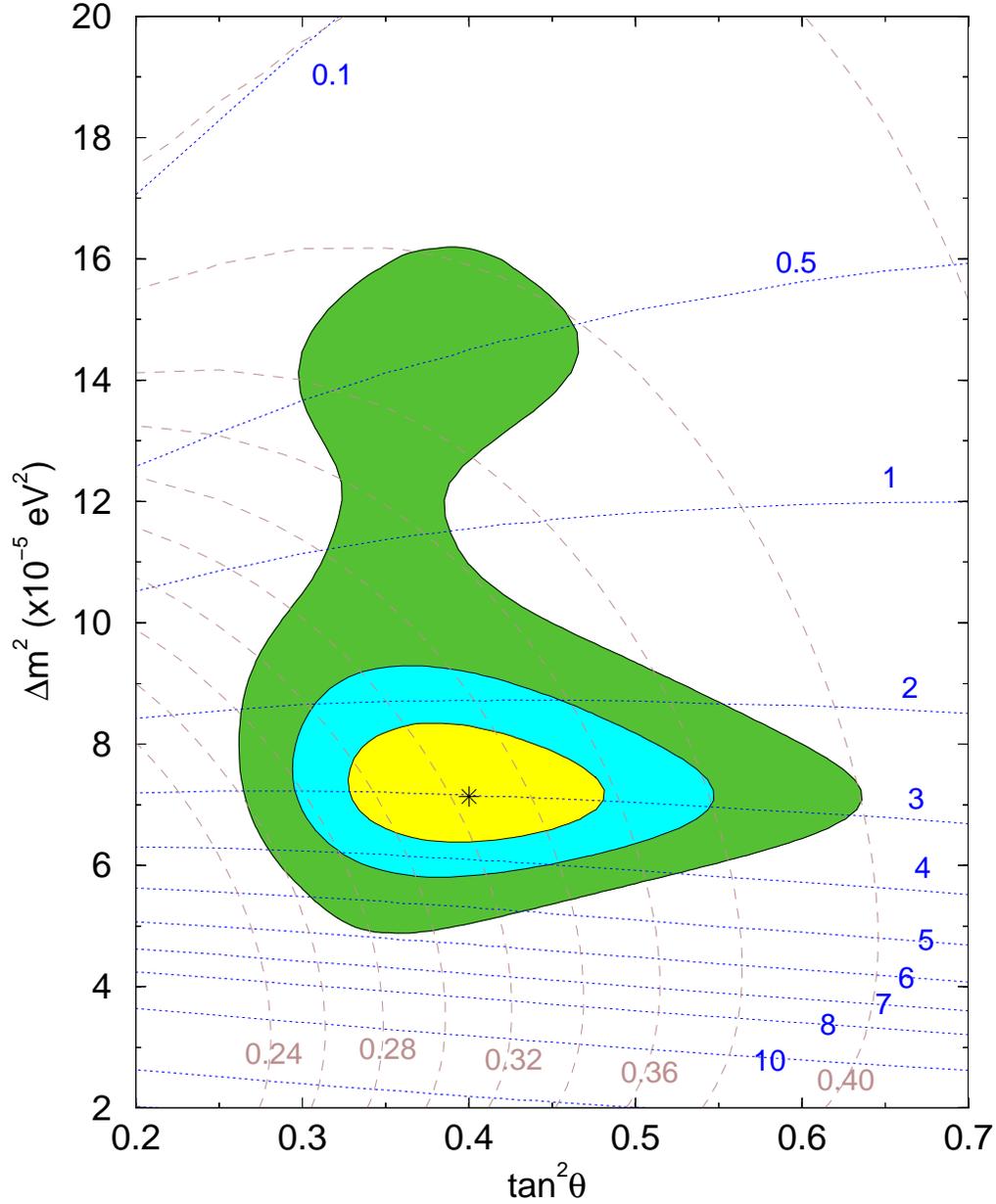}
\caption{Predictions for the CC/NC ratio and the Day-Night asymmetry at
SNO. The dashed lines are the lines of constant CC/NC ratio (numbers
at the curves) and the dotted lines show the lines of constant
$A_{DN}^{SNO}$ (numbers at the curves in \%).  We show also the
$1\sigma$ and $3\sigma$ allowed regions of the oscillation parameters
from the combined fit of the solar neutrino data and the KamLAND
spectrum. The best fit point is indicated by a star.}
\label{cc-nc}
\end{figure}

\noindent
{\it 2). Precise measurements of the {\rm CC/NC} ratio at SNO.}  In
fig.~\ref{cc-nc} we show the contours of constant CC/NC ratio with
finer grid than before.  We find predictions for the best fit point
and the $3\sigma$ interval:
\be
\frac{\rm CC}{\rm NC} = 0.32^{+ 0.08}_{-0.07} ~, ~~~ (3\sigma) . 
\ee

\noindent
{\it 3). The day-night asymmetry at SNO.}  In fig.~\ref{cc-nc} we show
also the contours of constant $A_{DN}^{SNO}$ for the energy threshold
5.5 MeV.  The best fit point prediction and the $3\sigma$ bound equal
\be
A_{ND}^{SNO} = 3.0 \pm 0.8 \% ~,~~~ (1 \sigma),~~~~~ A_{ND}^{SNO} < 6
\% ~~~~ (3 \sigma).
\label{dnas}
\ee

\begin{figure}[ht]
\centering\leavevmode
\epsfxsize=.8\hsize
\epsfbox{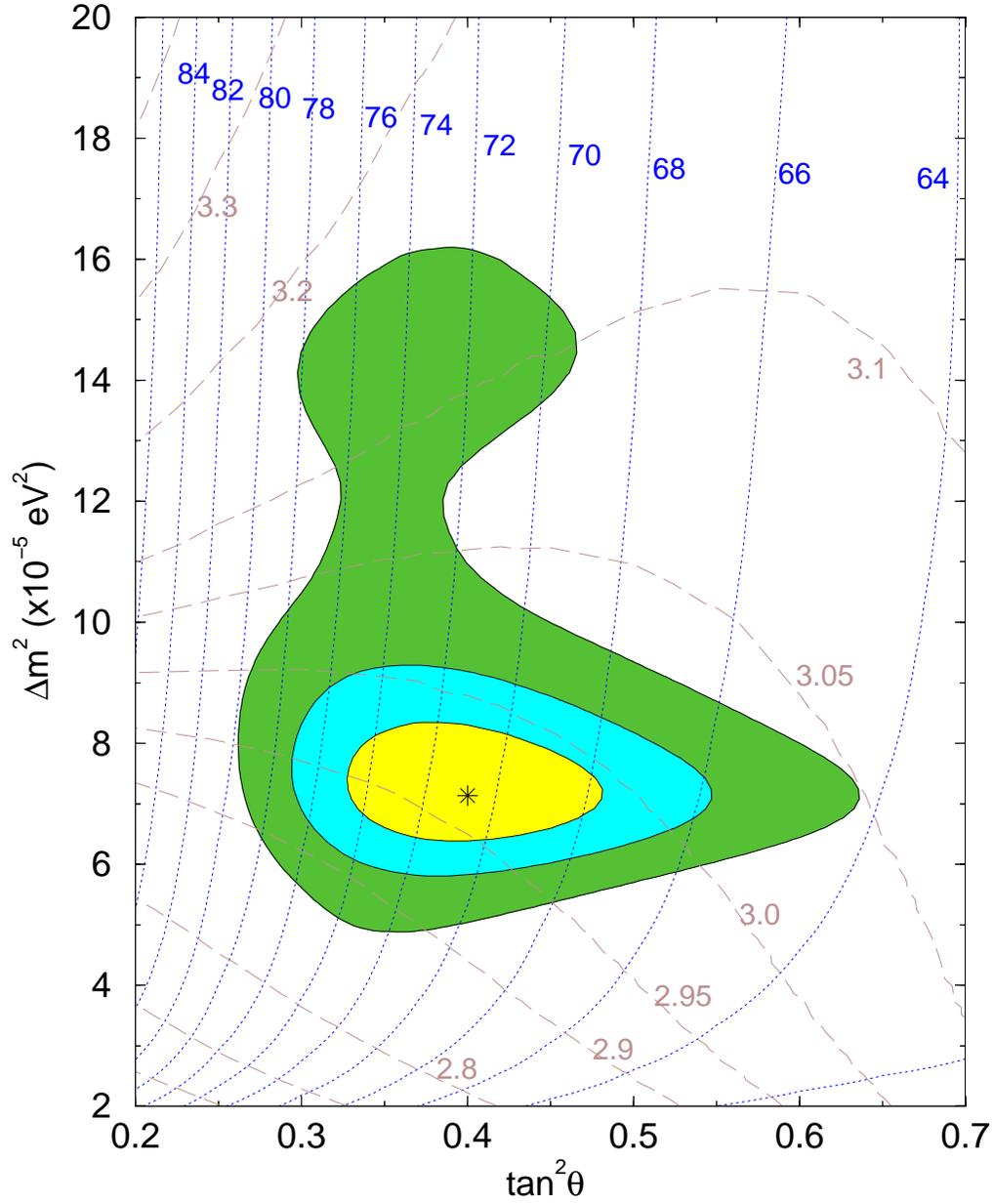}
\caption{
Predictions for the Germanium and Argon production rates.  The
dotted lines are the lines of constant Germanium production rate,
$Q_{Ge}$, and the dashed lines show the lines of constant Argon
production rate ,$Q_{Ar}$ (numbers at the curves in SNU), in the
$\Delta m^2 - \tan^2\theta$ plane. We show also the allowed regions of the
oscillation parameters from the combined fit of the solar neutrino
data and the KamLAND spectrum. The best fit point is indicated by a
star.}
\label{gal}
\end{figure}

\noindent
{\it 4). Germanium production rate. }  In the best fit point we
predict $Q_{Ge} = 71$ SNU.  We show in fig.~\ref{gal} the lines of
constant Ge production rate with finer (than before) grid.

For the Argon production we have $Q_{Ar} = 2.96$ SNU in the global
b.f. point. We show in fig.~\ref{gal} also the lines of
constant Argon production rate.

\section{Conclusions}

\noindent
1. The SNO-II fluxes have only slightly shifted the best fit point
toward smaller $\Delta m^2$ and $\tan^2 \theta$. The most important
improvements consist however of the stronger upper bounds on $\Delta
m^2$ and $\tan^2 \theta$. Those imply that the h-LMA region is
strongly disfavored and 1-2 mixing deviates substantially from the
maximal one:
\be
(\sin^2 \theta - 0.5) \sim \sin^2 \theta.
\ee
The 1-3 mixing does not change these results once the CHOOZ data are
included in the analysis.

\noindent
2. These improvements in measurements of the oscillation parameters
lead to a situation when physics of the solar neutrino conversion is
essentially (and quantitatively) determined. In the high energy part
of spectrum the averaged survival probability is close to the
non-oscillatory one.  For $E > 5$ MeV the effective $P$ is about 12\%
higher than $\sin^2\theta$. Oscillations give the dominant
contribution to this difference: $\sim 8 - 14 \%$, the rest is due to
the Earth regeneration effect. In spite of smallness of difference $(P
- \sin^2\theta)$ which is proportional to $\eta^2$, the depth of
oscillations is relatively large: $A_P \sim 0.45$ being proportional
to $\eta$. At low energies ($E < 1$ MeV), vacuum oscillations are the
dominant process with the matter corrections to $P$ below 5\%.

\noindent
3. After the SNO salt results the errors of determination of the
oscillation parameters become smaller than the values of parameters:
\be
\delta(\Delta m^2) < \Delta m^2, ~~~ \delta (\tan^2 \theta) < \tan^2 \theta. 
\ee 
This means that the solar neutrino studies enter a stage of precision
measurements.

For the forthcoming measurements we predict about 8 - 15\% upturn of
the energy spectrum at SNO.  If further measurements confirm the
absence of the upturn hinted by the present data, some physics
``beyond LMA" should be invoked.

The CC/NC ratio is expected to be $\approx 0.32$, the Day-Night
asymmetry: $A_{ND}^{SNO} \sim 3\%$ ($T_{eff} > 5.5$ MeV), the spectrum
shape parameter: $k = 1.0 +0.8/-0.5$.

\section{Acknowledgments}

One of the authors (P.C.H.) would like to thank FAPESP for financial
support.  The work of A.S. was supported by the TMR, EC-contract
No. HPRN-CT-2000-00148 and No. HPRN-CT-2000-00152.


\newpage

\newpage

\newpage

\newpage

\newpage

\newpage

\newpage

\newpage

\newpage

\end{document}